\documentclass[12pt,titlepage]{article}
\usepackage{amsmath}
\usepackage{amssymb}
\usepackage{graphicx}
\usepackage{caption2}
\usepackage{amsfonts}
\usepackage{cite}

\newcommand{\be}{\begin{equation}}
\newcommand{\ee}{\end{equation}}
\newcommand{\bea}{\begin{eqnarray}}
\newcommand{\eea}{\end{eqnarray}}
\newcommand{\bef}{\begin{figure}}
\newcommand{\ef}{\end{figure}}
\newcommand{\bt}{\begin{tabular}}
\newcommand{\et}{\end{tabular}}
\newcommand{\bno}{\begin{enumerate}}
\newcommand{\eno}{\end{enumerate}}

\begin{document}

\begin{center}

{\bf\large  On peaked solitary waves of Camassa-Holm equation}

\vspace{0.5cm}

Shijun Liao\\
\vspace{0.5cm}
State Key Laboratory of ocean Engineering, Dept. of Mathematics\\
School of Naval Architecture, Ocean and Civil Engineering\\
Shanghai Jiao Tong University, Shanghai 200240, China\\
\vspace{0.5cm}
(Email address:  sjliao@sjtu.edu.cn)

\end{center}

\hspace{-0.75cm} {\bf Abstract} {\em  Unlike the Boussinesq, KdV and BBM equations, the celebrated Casamma-Holm (CH) equation can model both phenomena of soliton interaction and wave breaking.  Especially, it has peaked solitary waves in case of $\omega=0$.  Besides,  in case of $\omega\neq 0$,  its solitary wave ``becomes  $C^\infty$ and there is no derivative discontinuity at its  peak'', as mentioned by Camassa and Holm  \cite{Camassa1993PRL}.  However,  it is found in this article  that the CH equation has peaked solitary waves even in case of $\omega\neq 0$.  Especially,  all of these peaked solitary waves  have an unusual property: their phase speeds have nothing to do with the height of peakons or anti-peakons.  Therefore,  in contrast to the traditional view-points,  the peaked solitary waves are a common property of the CH equation: in fact,  all  mainstream models of shallow water waves admit such kind of peaked solitary waves. }

\hspace{-0.75cm} {\bf PACS Number}: 47.35.Bb

\hspace{-0.75cm} {\bf Key Words}  Peaked solitary waves,  discontinuity, Camassa-Holm equation

\section{Introduction}

Since the solitary surface wave was discovered by John Scott Russell \cite{Russell1844} in 1834,   many models of solitary waves in shallow water  have been developed, such as the Boussinesq equation \cite{Boussinesq1872}, the  Korteweg  \&  de Vries (KdV)  equation \cite{KdV} and  the  Benjamin-Bona-Mahony   (BBM) equation \cite{Benjamin1972},  and so on.
The Boussinesq and KdV equations are integrable and can model the soliton interaction of solitary waves and propagating waves with permanent form.  However, they can not model breaking waves.   The  BBM equation has better analytic properties than the KdV equation, but it is not integrable, its traveling waves are not solitons and it can not model breaking waves (see \cite{Constantin2000}).   In contrast to the KdV, Boussinesq and BBM equations,  the celebrated Camassa-Holm (CH) equation \cite{Camassa1993PRL}
\begin{equation}
u_t+2\omega u_x-u_{xxt}+ 3 u u_x = 2 u_x u_{xx} + u u_{xxx}\label{geq:CH}
\end{equation}
can model  both  phenomena of soliton interaction and wave breaking (see \cite{Constantin2000}),  where $u(x,t)$ denotes the wave elevation, $x,t$ are the temporal and spatial variables, $\omega$ is a constant related to the critical shallow water wave speed,  the subscript denotes the partial differentiation, respectively.   Mathematically,  the CH equation is integrable and bi-Hamiltonian,
thus possesses an infinite number of conservation laws in involution \cite{Camassa1993PRL}.   In addition,
it is associated with the geodesic flow on the infinite dimensional Hilbert manifold of diffeomorphisms of line (see \cite{Constantin2000}).   Thus, the CH equation   (\ref{geq:CH})   has many intriguing physical  and  mathematical properties.   As pointed out by Fushssteiner \cite{Fushssteiner1996},  the CH equation (\ref{geq:CH}) ``has the potential to become the new master equation for shallow water wave theory''.

Especially, when $\omega=0$,  the CH equation (\ref{geq:CH}) has the peaked solitary wave \cite{Camassa1993PRL}
\[  u(x,t)  =  c \exp (-|x-c t|). \]
When $\omega \neq 0$,  ``the soliton solution of  (\ref{geq:CH})  becomes  $C^\infty$ and there is no derivative discontinuity at its  peak'', as mentioned by Camassa and Holm  \cite{Camassa1993PRL}.   As a result of it,   investigations of the peaked solitary waves of the CH equation (\ref{geq:CH}) were restricted in the case of $\omega=0$ (see \cite{Kalisch2005} for example).  

Currently,  the closed-form solutions of the peaked solitary waves of the KdV equation, the modified KdV equation, the Boussinesq equation and BBM equation are found by Liao \cite{Liao-arXiv-KdV}.    In this article,  it is found  that  the peaked solitary waves of the CH equation (\ref{geq:CH})  also  exist  even in case of $\omega \neq 0$.

\section{Peaked solitary waves of the CH equation in case of $\omega \neq 0$ }

Let us consider the propagating solitary waves of the CH equation   (\ref{geq:CH})  with permanent form.  Writing $\xi = x- c t$ and $w(\xi) = c \; u(x,t)$,  the original CH equation (\ref{geq:CH})   becomes
\begin{equation}
w''' - \left(1-\frac{2\omega}{c}\right)w' + 3 w w' = 2 w' w'' + w w''',  \label{geq:CH:2}
\end{equation}
subject to the boundary conditions
\begin{equation}
w\to 0, w'\to 0, w''\to 0, \;\; \mbox{as $|\xi|\to +\infty$}, \label{bc:CH:2}
\end{equation}
where the prime denotes the differentiation.

The linearized CH equation
\begin{equation}
w''' - \left(1-\frac{2\omega}{c}\right) w' =0 \label{geq:CH:Linear}
\end{equation}
has the peaked solitary wave
\begin{equation}
w(\xi) = A\; \exp(-\mu |\xi|),  \;\;  \mu = \sqrt{1-\frac{2\omega}{c}},
\end{equation}
provided $\omega < c/2$.

The solution of the equation (\ref{geq:CH:2})  can be expressed in the form
\[  w(\xi) = \sum_{n=1}^{+\infty} a_n \; \exp(-n \mu |\xi|), \]
where $a_n$ is a consant to be determined.   Write $A = w(0)$.  Using the homotopy analysis method (HAM) \cite{Liao1999, Liao1999JFM, LiaoBook2003, LiaoBook2012},  a analytic technique for highly nonlinear differential equations,  it is easy to gain the series solution
\begin{equation}
w(\xi) = w_0(\xi) +\sum_{m=0}^{+\infty} w_m(\xi).\label{series:w}
\end{equation}
Here
\[  w_0(\xi) = A \exp(-\mu |\xi|) \]
is the initial guess,  $w_m$ for $m\geq 1$ is governed by
\begin{equation}
{\cal L}\left[ w_m(\xi) -\chi_m \; w_{m-1}(\xi)  \right] = c_0 \; \delta_{m-1}(\xi),
\end{equation}
subject to the boundary condition
\begin{equation}
w_m(0)=0, \;  w_m\to 0, \;\;  \mbox{as $|\xi|\to +\infty$},
\end{equation}
where $c_0\neq 0 $ is an auxiliary parameter, called the convergence-control parameter,
\[  {\cal L}f = f''' -\mu^2 f' \]
is an auxiliary linear operator, and
\begin{eqnarray}
\delta_n &=&  w_n'''-\mu^2 w_n' + \sum_{j=0}^{n} \left[ 3 w_j w'_{n-j} -2w'_j w''_{n-j}-w_j w'''_{n-j} \right],\;\;\;\;\\
\chi_n &=& \left\{
\begin{array}{ll}
1 & \mbox{when $n>1$},\\
0 & \mbox{otherwise.}
\end{array}
\right.
\end{eqnarray}
The HAM is independent of small physical parameters.  Especially, unlike other analytic techniques, the HAM provides us a convenient way to guarantee the convergence of approximation series.  For details, please refer to Liao \cite{Liao1999, LiaoBook2003, LiaoBook2012}.  In fact, directly using the HAM-based mathematica package BVPh 1.0 (see Part II of \cite{LiaoBook2012}) for nonlinear boundary-value/eigenvalue problems,  it is straightforward to gain high-order  analytic approximations of   (\ref{geq:CH:2}) and (\ref{bc:CH:2}).    For details, please refer to the Appendix.

\begin{table}[t]
\begin{center}
\caption{$u'(0_+)$ of the $m$th-order analytic approximations and the corresponding residual squares of Camassa-Holm equation (\ref{geq:CH}) in case of $c=1, \omega=1/4$ given by the HAM-Based package BVPh 1.0  with the convergence-control parameter $c_0=-1$.  }

\vspace{0.5cm}

\begin{tabular}{|c|c|c|c|c|}
  \hline\hline
   &  \multicolumn{2}{c|}{$A=1/10$}  & \multicolumn{2}{c|}{$A=1/5$}  \\ \cline{2-5}
  $m$ &  $u'(0_+)$ & ${\cal E}_m$ & $u'(0_+)$ & ${\cal E}_m$ \\ \hline
  2 & -0.066733 & 1.1 E-9 & -0.123744 & 2.6E-7 \\
  \hline
  4 & -0.066668 & 1.6 E-13 & -0.122575 & 7.0E-10 \\
  \hline
  6 & -0.066667 & 2.2 E-16 & -0.122484 & 1.8E-11 \\
  \hline
  8 & -0.066667 & 2.9 E-19 & -0.122475 & 3.7E-13 \\
  \hline
  10 & -0.066667 & 2.8 E-22 & -0.122475 & 5.7E-15 \\
  \hline\hline
\end{tabular}
\label{Table:1}

\vspace{0.5cm}

\caption{$u'(0_+)$ of the $m$th-order analytic approximations and the corresponding residual squares of Camassa-Holm equation (\ref{geq:CH}) in case of $c=1, \omega=1/4$ given by the HAM-Based package BVPh 1.0 with the convergence-control parameter $c_0=-1$.  }

\vspace{0.5cm}

\begin{tabular}{|c|c|c|c|c|}
  \hline\hline
   &  \multicolumn{2}{c|}{$A=-1/10$}  & \multicolumn{2}{c|}{$A=-1/5$}  \\ \cline{2-5}
  $m$ &  $u'(0_+)$ & ${\cal E}_m$ & $u'(0_+)$ & ${\cal E}_m$ \\ \hline
  2 & 0.073804 & 1.3 E-9 & 0.152028 & 3.4E-7 \\
  \hline
  4 & 0.073854 & 1.7 E-13 & 0.152698 & 7.4E-10 \\ \hline
  6 & 0.073855 & 1.6 E-16 & 0.152748 & 9.3E-12 \\ \hline
  8 & 0.073855 & 2.2 E-19 & 0.152752 & 2.1E-13 \\ \hline
10  & 0.073855 & 2.2 E-22 & 0.152752 & 3.4E-15 \\
  \hline\hline
\end{tabular}
\label{Table:2}
\end{center}
\end{table}

 The accuracy of the $m$th-order approximation is defined by the averaged residual square
of (\ref{geq:CH:2}) in the domain $\xi\in[0,a]$:
\begin{equation}
{\cal E}_m = \frac{1}{a} \int_0^a \left[{\cal N}\left(\sum_{j=0}^{m}w_j\right) \right]^2 d\xi,
\end{equation}
where
\[ {\cal N}w = w''' - \mu^2 w' + 3 w w' - 2 w' w'' - w w'''. \]
Since the wave elevation decays exponentially, we use $a=10$ in this article.

\begin{figure}[thbp]
\centering
\includegraphics[scale=0.5]{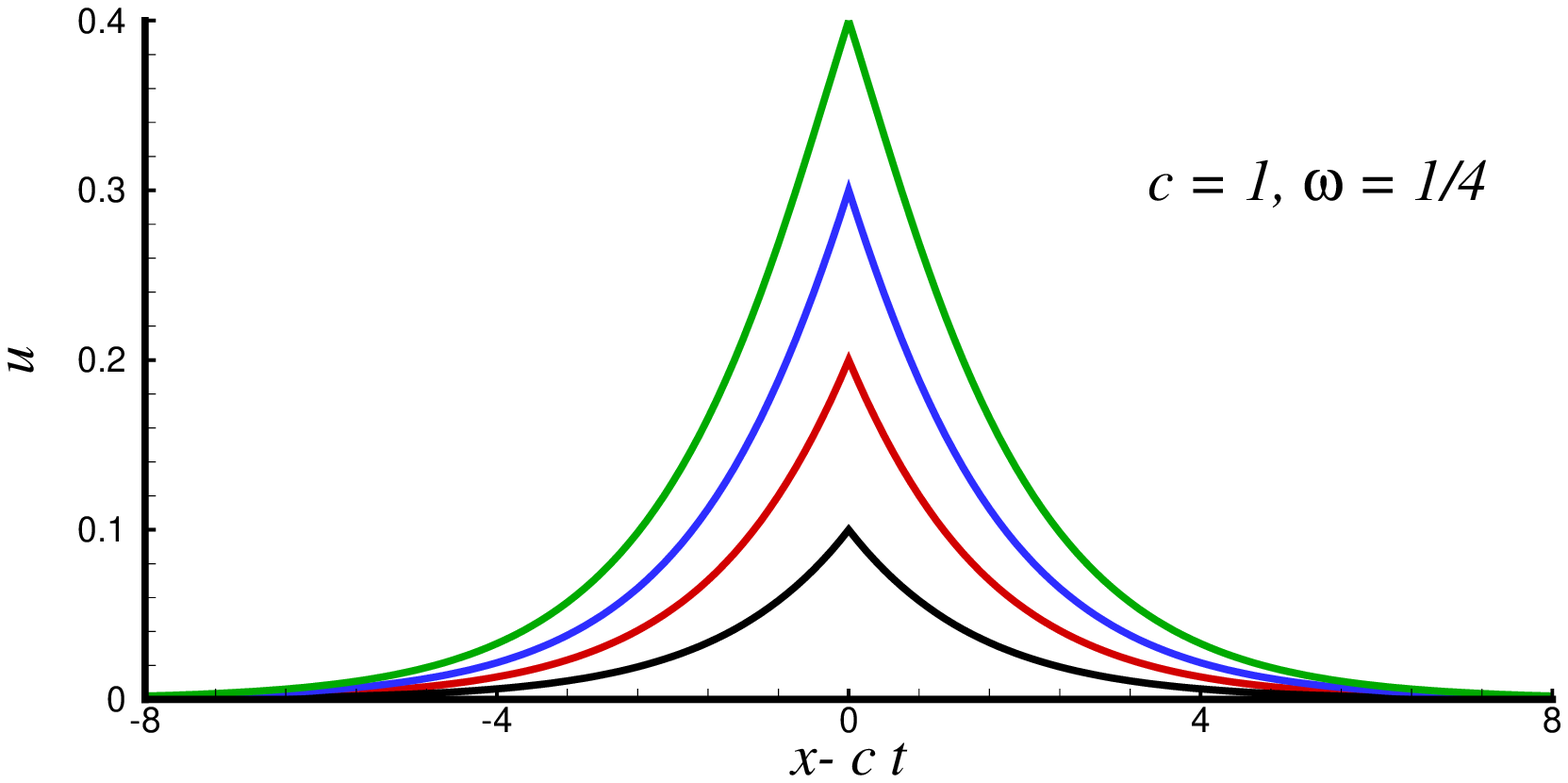}
\caption{ The peaked solitary waves $u(x,t)$ of the Camassa-Holm equation (\ref{geq:CH}) in case of $\omega=1/2$ with the same phase speed $c = 1$.  Black line: $A=1/10$;  Red line: $A=1/5$;  Blue line: $A=3/10$; Green line: $A =2/5$. }
\label{figure:omega0.25}

\centering
\includegraphics[scale=0.5]{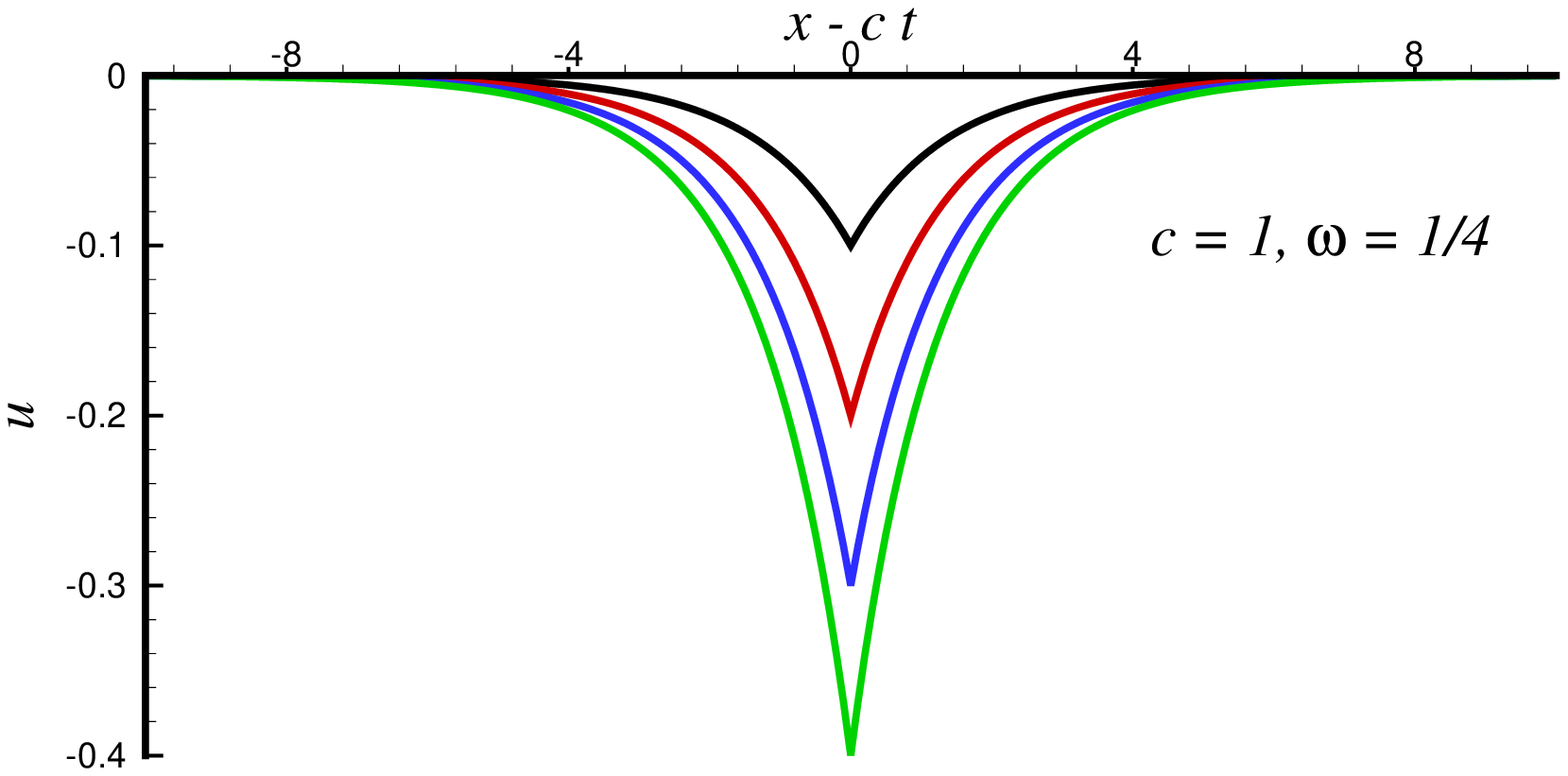}
\caption{ The peaked solitary waves $u(x,t)$ of the Camassa-Holm equation (\ref{geq:CH}) in case of $\omega=1/2$ with the same phase speed $c = 1$.  Black line: $A=-1/10$;  Red line: $A=-1/5$;  Blue line: $A=-3/10$; Green line: $A =-2/5$. }
\label{figure:omega-0.25}
\end{figure}

For simplicity, we  study the case $c = 1$ henceforth.  First,  let us consider  the case of $\omega = 1/4$.   As shown in Table~\ref{Table:1}, the averaged residual squares  of  the 10th-order analytic approximations decreases to 2.8$\times 10^{-22}$  in case of $A=1/10$ and to 5.7$\times 10^{-15}$ in case of $A=1/5$, respectively.   Besides,  the corresponding values of $u'(0_+)$ (the limit is taken as $\xi \to 0$ from the right) quickly  converge to -0.066667 and -0.122475, respectively.   Similarly,   as shown in Table~\ref{Table:2}, the averaged residual squares  of  the 10th-order analytic approximations decreases to 2.2$\times 10^{-22}$  in case of $A=-1/10$ and to 3.4$\times 10^{-15}$ in case of $A=-1/5$, respectively.  These averaged residual squares are much smaller than those of numerical ones (see  \cite{Kalisch2005}  as an example).      Thus, without doubt,  they are the solutions of the CH equation (\ref{geq:CH}) in case of $\omega=1/4$.      It is very interesting that all of the corresponding wave elevations are peaked solitary waves,  as shown in Fig.~\ref{figure:omega0.25} with peakons  and Fig.~\ref{figure:omega-0.25}  with anti-peakons for different values of  $A$.   Thus,   the peaked solitary waves of the CH equation (\ref{geq:CH})  indeed  exists even in case of $\omega \neq 0$.     Here, as pointed out by  Constantin and Molinet \cite{Constantin2000-B},  all of these peaked solitary waves  should be understood mathematically as weak solutions of the CH equation (\ref{geq:CH}).  However, physically,  this  kind of discontinuity of wave elevation widely appears in fluid mechanics, such as dam break \cite{Zoppou2000AMM}  in hydrodynamics and shock waves in aerodynamics, which have clearing physical meanings.   In fact,  such kind of discontinuous problems belong to the so-called Riemann problem \cite{Bernetti2008JCP, Rosatti2010JCP, Wu2008IJNMF, Zoppou2000AMM},  a classical field of fluid mechanics.

\begin{figure}[thbp]
\centering
\includegraphics[scale=0.5]{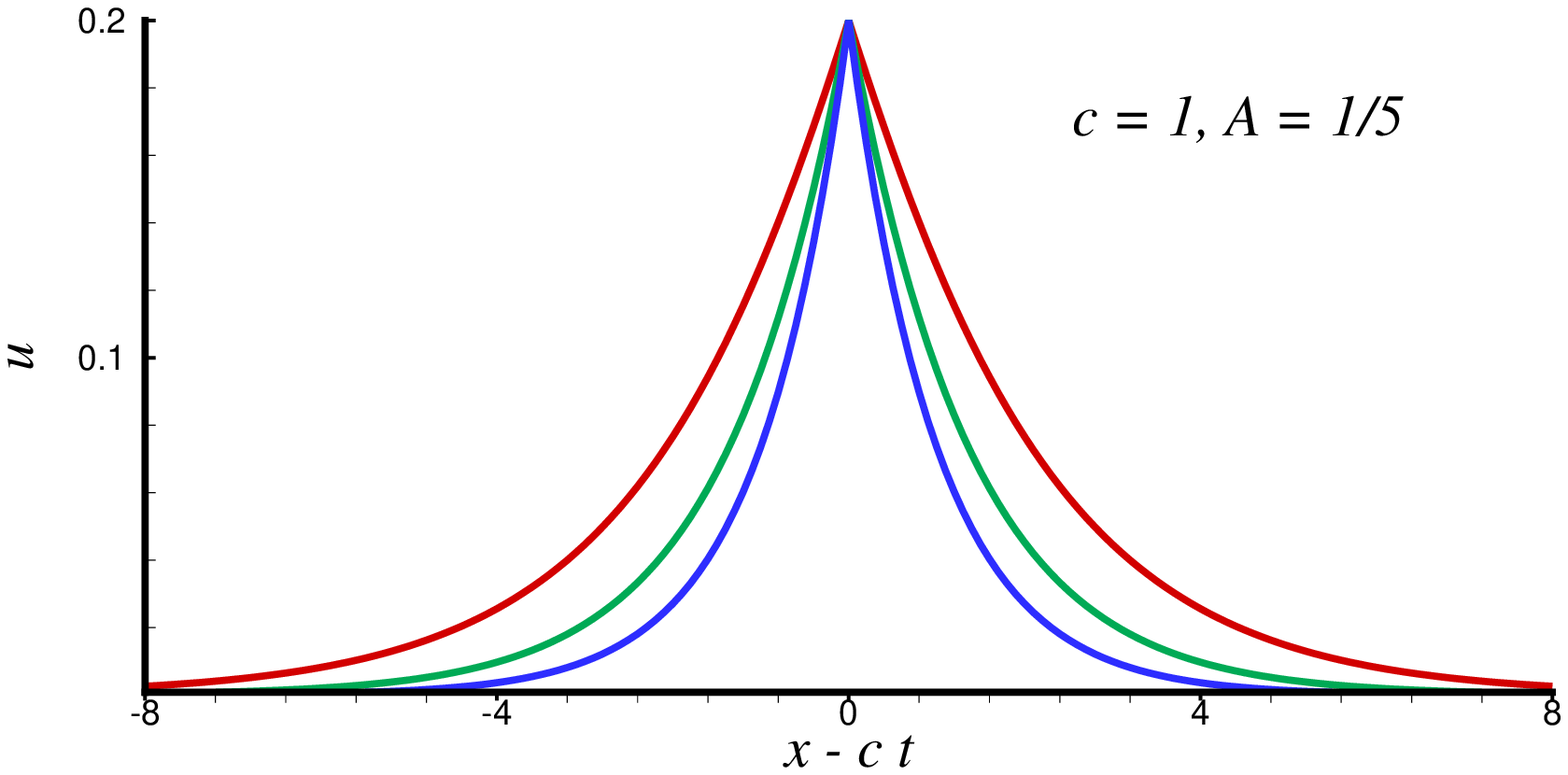}
\caption{ The peaked solitary waves $u(x,t)$ of the Camassa-Holm equation (\ref{geq:CH}) in case of $A=1/5$ with the same phase speed $c = 1$.  Red line: $\omega=1/3$;  Green line: $\omega=1/5$;  Blue line: $\omega = 0$. }
\label{figure:A0.2}
\end{figure}

\begin{figure}[thbp]
\centering
\includegraphics[scale=0.5]{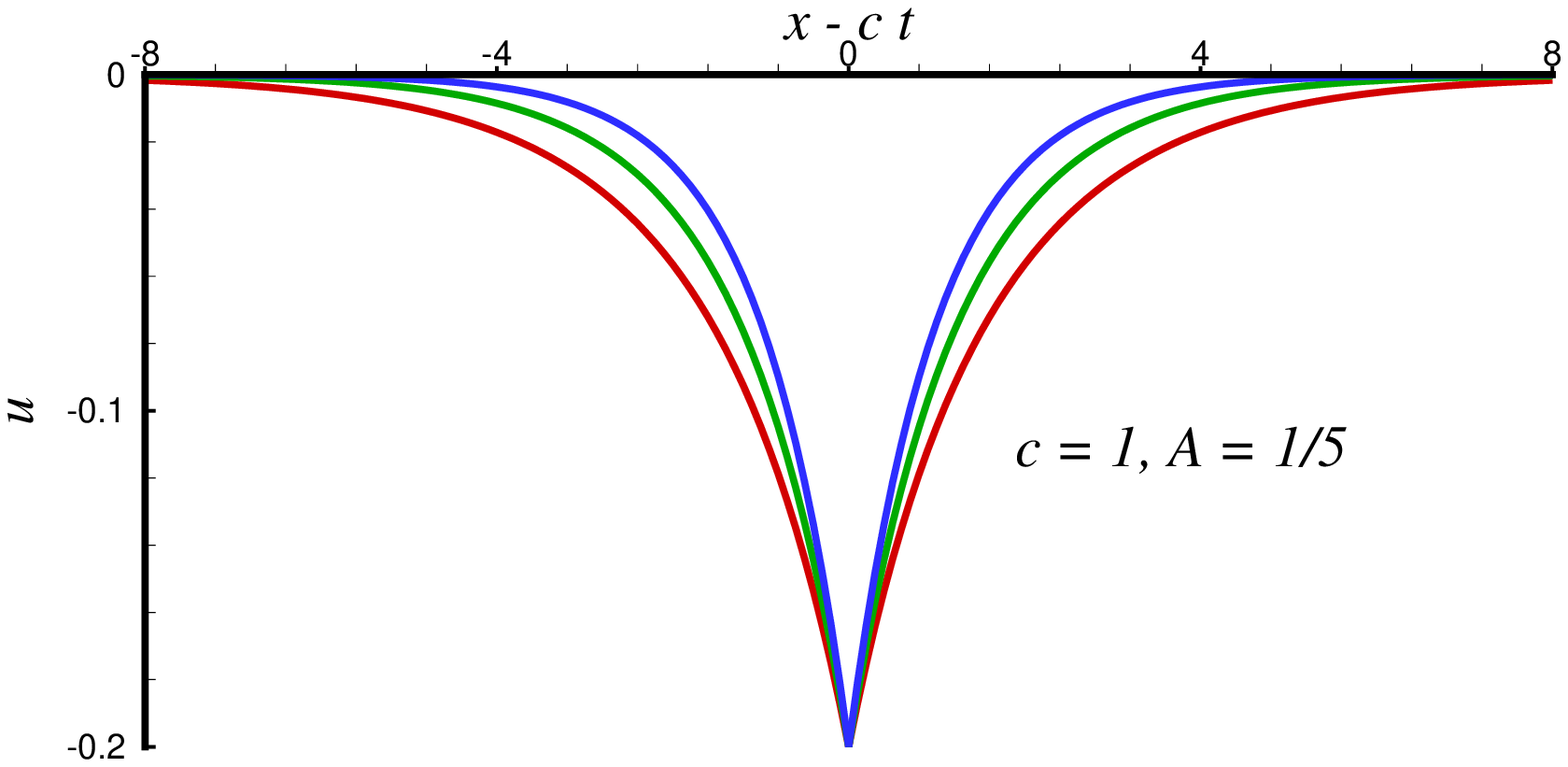}
\caption{ The peaked solitary waves $u(x,t)$ of the Camassa-Holm equation (\ref{geq:CH}) in case of $A=-1/5$ with the same phase speed $c = 1$.  Red line: $\omega=1/3$;  Green line: $\omega=1/5$;  Blue line: $\omega = 0$. }
\label{figure:A-0.2}
\end{figure}

Secondly, let us consider the case of $c=1$ and $A = \pm 1/5$ but different values of $\omega$.  Similarly, using the HAM-based mathematica package BVPh 1.0 \cite{LiaoBook2012},  it is straightforward to gain the corresponding convergent analytic approximations with high accuracy.  It is found that all of them have peakons (when $A=1/5$) or anti-peakons (when $A=-1/5$), as show in Fig.~\ref{figure:A0.2} and Fig.~\ref{figure:A-0.2}, respectively.    Note that, when $\omega=0$, the BVPh 1.0 gives the closed-form solution  $w = A \exp(-|x-c t|)$ of the peaked solitary wave, reported by Casamma and Holm \cite{Camassa1993PRL}.  It is found that,  as $\omega$ becomes larger, the wave elevation decays more slowly.  This confirms once again that the CH equation (\ref{geq:CH})  possesses the peaked solitary waves even when $\omega\neq 0$.

\section{Concluding remarks}

As mentioned by many authors, the CH equation (\ref{geq:CH}) has many intriguing  properties in physics and mathematics.   Physically,    it can model  both phenomena of soliton interaction and wave breaking.   Mathematically,  it is  integrable and bi-Hamiltonian,
thus possesses an infinite number of conservation laws in involution \cite{Camassa1993PRL}, and besides it is associated with the geodesic flow on the infinite dimensional Hilbert manifold of diffeomorphisms of line.     In addition, like Euler equation whose limiting wave also has a crest with a corner \cite{Stokes1894},  the CH equation admits peaked traveling  waves not only for $\omega=0$ but also for $\omega\neq 0$, as found in this letter.    Especially, the phase speed of these peaked solitary  waves has nothing to do with  the height of the peakon or anti-peakon:  this is an unusual  property of the peaked  solitary waves.

Note that the closed-form solutions of peaked solitary waves of the KdV equation, Boussinesq equation and BBM equation have been found by Liao \cite{Liao-arXiv-KdV}.    Therefore,   peaked  waves seem to be common for the mainstream models of shallow water waves.   Indeed, the elevation of these peaked solitary waves have discontinuity at crest.  However,  this kind of discontinuity widely appears in fluid mechanics such as dam break and shock waves which have clear physical meanings.   In fact, such kind of discontinuous problems belong to the classic Riemann problems.  Therefore, the discontinuity of these peaked solitary waves are acceptable not only in mathematics but also in physics.   It is true that such kind of peaked solitary waves have never been observed.   So,  it is an interesting and  challenging  work  to  observe them in laboratory.    Certainly,  further theoretical, numerical and experimental  investigations on these peaked waves are necessary.

Finally,  it should be emphasized that,  the discontinuity and/or  singularity exist widely in natural phenomena, such as dam break in hydrodynamics, shock waves in aerodynamics,  black holes in general relativity equation  and so on.  Indeed, the discontinuity and/or singularity are difficult to handle by traditional methods.  But, the discontinuity and/or singularity  can greatly enrich our understandings about the real world, and therefore should not be  evaded.     

\section*{Acknowledgement}

This work is partly supported by the State Key Lab of Ocean Engineering (Approval No. GKZD010056-6) and the National Natural Science Foundation of China.

\pagebreak\newpage

\bibliographystyle{unsrt}

\pagebreak\newpage

\begin{center}
{\bf Appendix\\ 
\vspace{0.5cm} 
 The use of HAM-based Mathematica package BVPh 1.0}
\end{center}

Based on the HAM [9-11], the mathematica package BVPh 1.0 for nonlinear boundary-value/eigenvalue problems is developed and issued by Liao (Part II) [11], which is free available online.
Using the BVPh 1.0, it is straightforward to gain the analytic approximations of the peaked solitary waves of the equations (2) and (3) for given $w(0)=A$.  Here, we briefly describe how to do it in case of $c = 1, \omega=1/4$ and $A=1/10$.

\begin{enumerate}
  \item First, download the  BVPh 1.0 (the code file is named by BVPh\_1.0.txt) online ( http://numericaltank.sjtu.edu.cn/BVPh.htm )
       and save it in a directory such as C:/math/CH as an example.

  \item Then, run the computer algebra system Mathematica, and type the following command one by one:
\end{enumerate}
{\small
\begin{verbatim}
SetDirectory["C:\math\CH"];

<<InputCH.txt
\end{verbatim}
}

The file named InputCH.txt contains the following Mathematica commands and necessary definitions for BVPh 1.0:
{\small
\begin{verbatim}
(* Install the BVPh 1.0  *)
<<BVPh1_0.txt;

(* Define the physical and control parameters *)
TypeEQ      =  1;
ApproxQ     =  0;
ErrReq      =  10^(-30);
zRintegral  =  10;

(* Define the governing equation *)
mu2 = 1-2*omega/c;
f[z_,u_,lambda_] := D[u,{z,3}]-mu2*D[u,z]  \
                  + 3*u*D[u,z] - 2*D[u,z]*D[u,{z,2}] - u*D[u,{z,3}] ;

(* Define Boundary conditions *)
zR  = Infinity;
OrderEQ  = 3;
BC[1,z_,u_,lambda_] := Limit[u-A, z -> 0 ];
BC[2,z_,u_,lambda_] := Limit[u, z -> zR ];
BC[3,z_,u_,lambda_] := Limit[D[u,z], z -> zR ];

(* Define initial guess *)
mu = Sqrt[mu2];
u[0]  = A*Exp[-mu*z];

(* Define output term *)
output[z_,u_,k_]:= Print["output = ",D[u[k],z] /. z->0//N];

(* Defines the auxiliary linear operator *)
L[u_] := D[u,{z,3}] - mu2 * D[u,z];

(*  Print input and control parameters  *)
PrintInput[u[z]];

(* Set convergence-control parameter c0 and physical parameters *)
c0    =  -1 ;
A     =  1/10;
omega =  1/4;
c     =   1;
Print[" c0  =  ",c0, "  omega  =  ",omega, "  c   = ", c, "  A  = ",A];

(*  Gain up to 10th-order HAM approximation *)
BVPh[1,10];

(*  Get results in the whole domain  *)
For[k=0,k<=10,k++,W[k] = U[k] /. z-> Abs[x]];

(* Show the 5th and 10th-order approximation  *)
Plot[{W[5],W[10]},{x,-10,10},PlotRange->{Min[A,0],Max[A,0]}]
\end{verbatim}
}

\end{document}